\documentstyle[sprocl]{article}
\def\Journal#1#2#3#4{{#1} {\bf #2}, #3 (#4)}

\def\PRD{{\em Phys. Rev.} D}
\begin{document}

\title{ FORCED KINK OF $\lambda \phi ^4$ MODEL}

\author{Ya.M. SHNIR{\footnote{On leave from  
Institute of Physics, Minsk, 
Belarus}}}

\address{DAMTP, University of Cambridge, Silver Street, Cambridge CB3 9EW}

\maketitle\abstracts{
The motion of a one-dimensional kink and  its energy losses are  
considered as a model 
of interaction of nontrivial topological field configurations with  
external 
fields.}
 
Let us consider the 2D $\lambda \phi^4$ 
theory with an additional linear perturbation
describing by the Lagrangian 
\begin{equation}                                    \label{Lagrang} 
L = \frac{1}{2} {\dot {\phi}}^2 - \frac{1}{2}{{\phi}'}^2 - \frac{\lambda} 
{4}\left(\phi ^2 - \frac{m^2}{\lambda}\right)^2 - \varepsilon   
\frac {m^3}{\sqrt \lambda} \phi,   
\end{equation} 
where the dimensionless parameter $\varepsilon \ll 1$. 
In order to solve corresponding  field equation
we can use an expansion in powers of $\varepsilon$: 
$\phi = \phi_0 + \varepsilon \phi_1 + \varepsilon ^2 \phi _2 + \dots $.  
The zero-order approximation gives the classical equation  
\begin{equation}                                             \label{zero-equ} 
{\ddot {\phi _0}} - {\phi _0}'' - m^2 \phi _0 + \lambda \phi _0 ^3 = 0, 
\end{equation} 
with the well known kink solution \cite{AdDash}  
$ 
\phi _0 = \frac{m}{\sqrt {\lambda}} ~{\rm {th}}~ \frac {mx}{\sqrt 2} 
$. The first order corrections to this  solution can be obtained from 
the next equation 
\begin{equation}                   \label{first-equ} 
\frac {d^2}{d t^2} \phi _1 + D^2 \phi _1 +  \frac {m^3}{\sqrt \lambda}  = 0, 
\end{equation} 
where the operator $D^2 = - \frac {d^2}{d x^2} - m^2 + 3m^2 ~{\rm {th}}^2~ \frac {mx}{\sqrt 2} 
$. 

 In order to find the corrections to the kink solution we can use the expansion  
of  $\phi _1$ on the normalizable eigenfunctions 
$\eta _n (x)$ of the operator $D^2$ which describe the scalar field  
fluctuations on the kink background, i.e. one can write
$\phi _1 = \sum\limits_{n=0}^{\infty} C_n(t) \eta _n (x)$.

If we substitute the expansion $\phi _1$ into eq.(\ref{first-equ}) 
we obtain 
\begin{equation}                   \label{first-equ-exp} 
\sum_{n=0}^{\infty}  
\biggl({\ddot C}_n(t) + \omega _n ^2 C_n(t)\biggr) \eta _n (x)    +  \frac {m^3}{\sqrt \lambda}  = 0 
\end{equation} 
 
Using the orthogonality relations one can make a projection of   
eq.(\ref{first-equ-exp}) onto the modes $\eta _n (z)$ that allows to define the time
dependent functions $C_n(t)$. Collecting the contributions from all modes we find
the first order correction to the kink  
configuration: 
\begin{equation}               \label{phi-one}             
\phi _1 = \frac {m}{\sqrt\lambda} \left\{ - \frac {3}{4} m^2 t^2 
{\eta }_0  - \frac {1}{4 } {\eta }_{k{_0}}   
+ a_1 e^{i \omega _1 t} \frac {~{\rm {sh}}~z}{{\rm {ch}}^2~z} +  
\sum _{k=0}^{\infty} a_k {\widetilde C}_k(t) \eta_k (x)\right\}  
\end{equation} 
where $z=mx/\sqrt 2$, ${\widetilde C}_k(t) = e^{i \omega _k t}$ and ${\eta }_{k{_0}} =  
3 ~{\rm {th}}^2~z - 1 $. 
  
The first term, connected with the zero mode contribution,  
describes the motion of the kink with a constant acceleration $a = -\varepsilon
\frac{3m}{\sqrt 2} = F/M,$ where $F=-2\varepsilon\frac{m^4}{\lambda}$ is the external
force and the kink mass is $M=2\frac{{\sqrt 2} m^3}{3\lambda}$.  
 The second term corresponds to a shift of the vacuum value of the scalar  
field \cite{KisSh}. 
  
The expression (\ref{phi-one})  allows to calculate the first order corrections  
to the kink energy ${\cal E}$. Substituting  
$\phi = \phi_0 + \varepsilon \phi_1$, we have,
as one could expect,  
$
{\cal E} = M + \varepsilon^2 \int\limits_{-\infty}^{\infty} dx  
\displaystyle  \frac{1}{2}  
{\dot {\phi_1}}^2 = M + \displaystyle \frac {M V^2}{2},
$   
where $V = \varepsilon 3mt/\sqrt 2 = at$ is the kink velocity.  
Note, that the changing of the kink kinetic energy is equal to the changing of  
the potential energy of the field due to linear perturbation.

The second order correction  $\phi_2$ to the kink solution $\phi _0$ 
 can be found from the next equation:  
\begin{equation}                   \label{second-equ} 
\biggl(\frac {d^2}{d t^2} + D^2\biggr) 
 \phi _2 + 3  \lambda \phi _0 \phi_1^2 = 0, 
\end{equation} 
Suppose that at the moment $t=0$ all oscillation modes are excited. Using again the  
expansion of   
$\phi _2$ on the eigenfunction 
$\eta _n (x)$ of the operator $D^2$ we write 
$
\phi _2 = \sum\limits_{n=0}^{\infty} \alpha_n(t) \eta _n (x).
$
Substituting this expansion in   
equation (\ref{second-equ}) 
after the projection onto the zero mode one can obtain: 
$$
\frac {4 \sqrt 2}{3m} {\ddot \alpha}_0 + 3  \lambda  
\int dx 
~\phi_0 \phi_1 ^2 \eta _0 = 0, $$ 
where $\phi _1$ is defined by eq.(\ref{phi-one}). 
The solution of this equation gives the next expression for the second order
correction to the kink position:
{\small
\begin{eqnarray}           \label{kink-corr-zero-2} 
\delta x^{(2)} = \varepsilon ^2 \frac{{\sqrt{2 \lambda }}}{m^2} \alpha_0 = 
- \varepsilon ^2 \frac {9 \pi }{4 {\sqrt 2}m} \biggl[\displaystyle 
 \sum_{k,k'=0}^{\infty}\sin (\omega_k - \omega_k') t ~\frac  
{\omega _k^2 -  
\omega _{k'}^2}{ (\omega _k - \omega _k')^2} \displaystyle \frac {2 +  
\frac {k^2 + {k'}^2}{2}} 
{~{\rm sh}\ \frac{\pi (k+ k')}{2}}  
\nonumber\\ -  \frac{1}{12} \left( 
\omega_1^2 t^2 - \frac{11}{2}\right) \cos \omega_1 t - 
 \sum_{k=0}^{\infty}\sin  \omega_k t \frac {k}{{\rm sh} 
\frac{\pi k}{2}} ~\left\{1 +  \left( -\omega _k ^2 t^2   
 + 6 + \frac {\omega _k^2}{m^2}\right)\frac{k}{4} \right\}
\nonumber\\
+  \sum_{k=0}^{\infty} \cos (\omega_k - \omega_1)t~ 
\frac{1 + k^2}{{\rm ch} \frac{\pi k}{2}} (\omega_k + \omega_1)^2     
+  \sum _{k=0}^{\infty} \omega_k t \cos \omega_k t ~ \frac {k^2}{{\rm sh} 
\frac{\pi k}{2}} + \frac{1}{3} \omega_1 t \sin \omega_1 t \biggr].\nonumber 
\end{eqnarray}  
 }
All these terms correspond to the oscillations of the kink by interaction with  
the vibrational modes. 
The energy of the kink interaction with phonons can be  
calculated from the second order correction $\delta x^{(2)} $  
eq.(\ref{kink-corr-zero-2}). Indeed, this correction is 
a sum of oscillations 
with the frequencies $\omega_k$ and different amplitudes $\delta x_k$.  
Thus, for the large time interval $t \gg 1$ 
the energy of each oscillation can be written as  
\begin{eqnarray} 
E_k = M\frac{\delta {x_k^{(2)}}^2\omega_k^2}{2} \approx M\frac{\varepsilon^4 3^4 
\pi^2}{2^{13}} m^4t^4 \frac {k^4(4+k^2)^3}{~{\rm sh}^2 
\frac{\pi k}{2}} = MV^4 \frac{\pi^2}{2^{11}}\frac {k^4(4+k^2)^3}{~{\rm sh}^2 
\frac{\pi k}{2}}, \nonumber
\end{eqnarray} 
Introducing the integration over momenta $k$ instead of  
sum one can estimate  the total  
energy of interaction between the kink and phonons as   
$
\delta E^{(2)}  \approx 3 \pi MV^4  
$,
that is much more than the second order relativistic correction.

In a simular way the second order corrections to the 
other kink modes can be obtained. Here we consider these 
corrections supposing that at the moment $t=0$ all oscillation modes are  not 
excited and take the limit $m t \gg 1$. In this case
the correction of the second order to the kink classical field is
 $$ 
\phi _2 = \frac {9 m}{2^8 \sqrt\lambda} m^4 t^4 \left\{ 3 \pi \eta_1 +  
i \sum _{k\not= 0} \frac{k^2}{1+k^2}~ \frac {1}{{\rm sh}\frac{\pi k}{2}} ~\eta_k 
\right\}. 
$$
The next order correction to the energy is
$$
\delta^{(4)} {\cal E} = \varepsilon^4 \int dx   
\frac{1}{2} {\dot {\phi_2}}^2 = \frac{3}{ 2^{10}} \left( 3 \pi^2 - 
\frac{1}{4} \int dk~ \frac{4+k^2}{1+k^2}~ \frac{k^4}{{\rm sh}^2\frac{\pi k}{2}} 
\right) M V^4 m^2 t^2. 
$$
The first term in the roots corresponds to the kink mass changing due to  
its bounding with the $\eta_1$ mode.      
Numerical calculation of the integral over $k$ in the roots gives the  
value $\approx 5.406$, i.e.   
the second term is, in two order,  
smaller then first one.  
This term corresponds to the correction to energy of the configuration connected 
with the continuum modes excitation.  
 
More detailed description of this problem as well as 
the application of perturbative approach to the description of the problem of
interaction between 't Hooft-Polyakov monopole and an external uniform magnetic field
one can found in paper \cite{KisSh}   
 

\section*{References}

\begin{thebibliography}{99}

\bibitem{AdDash} 
R. Dashen {\it et al.},  \Journal{\PRD}{D10}{4136}{1975}.
\bibitem{KisSh}  V.G. Kiselev and Ya.M. Shnir, Preprint ICTP IC/97/128 (1997).

\end{thebibliography}
\end{document}